\documentclass[12pt]{article}

\usepackage{preprintstyle}
\usepackage[utf8]{inputenc}

\newcommand{\beq}{\begin{equation}}
\newcommand{\eeq}{\end{equation}}
\newcommand{\bal}{\begin{aligned}}
\newcommand{\eal}{\end{aligned}}
\newcommand{\rmd}{\mathrm d}

\usepackage{physics, xparse}
\usepackage[normalem]{ulem}
\usepackage{cancel,xcolor}

\title{Extremal Black Hole Decay\\ in de Sitter Space}

\author{Lars Aalsma$^{a}$,}
\author{Jan Pieter van der Schaar$^{b}$,}
\author{Manus R. Visser$^{c}$,}
\emailAdd{laalsma@asu.edu}
\emailAdd{j.p.vanderschaar@uva.nl}
\emailAdd{mv551@cam.ac.uk}
\affiliation{$^a$Department of Physics and Beyond: Center for Fundamental Concepts in Science, Arizona State University, Tempe, Arizona 85287, USA}
\affiliation{$^b$Delta Institute for Theoretical Physics, Science Park 904, PO Box 94485, 1090 GL Amsterdam, the
Netherlands}
\affiliation{$^c$Department of Applied Mathematics and Theoretical Physics, University of Cambridge, Wilberforce Road, Cambridge CB3 0WA, United Kingdom}

\abstract{The decay of extremal charged black holes has been a useful guidance to derive consistency conditions in quantum gravity. In de Sitter space it has been argued that requiring (extremal) charged Nariai black holes to decay without forming a big crunch singularity yields the Festina Lente (FL) bound: particles with mass $m_s$ and charge $q$ should satisfy $m_s^2 \gg M_pHq$, where $M_p$ is the Planck mass and $H$ the Hubble parameter. Using a tunneling approach we show that the decay probability of charged black holes in de Sitter space in the s-wave sector is  $P\sim \exp(\Delta S_b)$, where~$\Delta S_b$ is the change in the black hole entropy. We find that the FL bound corresponds to $\Delta S_b \leq -1$ in the Nariai and probe limit. However, taking into account backreaction we identify  unsuppressed decay channels, which might be subdominant, that violate this bound but nonetheless do not result in a big crunch for every observer.}

\begin{document}

\maketitle

\section{Introduction}
Consistency principles have proven to be invaluable tools in uncovering universal properties of quantum gravity. This method, which does not assume a specific quantum gravity theory, has yielded a plethora of constraints that effective theories must satisfy in order to be consistent with quantum gravity \cite{Brennan:2017rbf,Palti:2019pca,vanBeest:2021lhn,Harlow:2022ich}. Nowadays, the derivation of quantum gravity constraints on effective theories is a central aspect of the swampland program, although various principles were established prior to the formalization of this concept, see e.g. \cite{Banks:1988yz,Giddings:1988cx,Abbott:1989jw,Kallosh:1995hi}.

Black holes play a central role in this framework with many constraints being derived from, or tied to, properties of (extremal) black hole solutions. For instance, black hole evaporation leads to breaking of global symmetries, suggesting the constraint that all global symmetries in quantum gravity must either be broken or gauged \cite{Banks:1988yz,Banks:2010zn}. Additionally, the Weak Gravity Conjecture (WGC) is a consequence of the requirement that extremal charged black holes should be able to decay \cite{Arkani-Hamed:2006emk}.   In its simplest form, the WGC states that---to be compatible with quantum gravity---effective theories with a U(1) gauge field should contain a superextremal massive, charged state. For four-dimensional theories the explicit bound is $m \leq q/(2\sqrt{\pi G})$ in ADM units (for the definition see Sec. \eqref{sec:flatRN}). This allows extremal black holes to decay via Schwinger pair production. There exists substantial and convincing evidence supporting the WGC, as reviewed for instance in \cite{Harlow:2022ich}.

Still, the precise mechanism responsible for breaking global symmetries has remained somewhat elusive although wormholes seem  to play an important role~\cite{Giddings:1988cx,Abbott:1989jw,Kallosh:1995hi,Belin:2020jxr,Bah:2022uyz}. Recent works have suggested a mechanism of global symmetry breaking that involves replica wormholes and relates global symmetry breaking to unitary black hole evaporation \cite{Harlow:2020bee,Chen:2020ojn,Hsin:2020mfa,Milekhin:2021lmq}. This is an important insight that allows for a more quantitative understanding of the magnitude of global symmetry breaking during black hole evaporation. Yet, computations have mostly been performed in two-dimensional theories and it is an open question how these ideas can be applied to higher-dimensional theories. 
There are also open questions related to the WGC.   An unambiguous formulation of the WGC exists only for extremal black holes in flat space. In anti-de Sitter space the correct version of the WGC that takes into account negative curvature remains a subject of ongoing study and various proposals have been put forward~\cite{Nakayama:2015hga,Montero:2018fns,Aharony:2021mpc,Palti:2022unw,Andriolo:2022hax,Cho:2023koe}. 

The present paper is centered around extremal black hole decay in de Sitter space. To understand the analogue of the WGC in de Sitter space, a new phenomenon needs to be taken into account: the existence of a black hole of maximum size at fixed charge (that fits within the cosmological horizon). In \cite{Montero:2019ekk} Montero, Van Riet and Venken argued that these solutions, known as charged Nariai black holes, have a potential instability under emission of light, charged particles. By taking a near-horizon limit in which decays are governed by Schwinger pair production, \cite{Montero:2019ekk} gave evidence that such decays lead to a big crunch singularity and should therefore be prohibited. This led to the Festina Lente (FL) bound  $m_s^2 \gg  M_p H q$, where $M_p$ the Planck mass, $H$ is the Hubble parameter,  $q$ the charge of the particle, and $m_s$ is the appropriate definition of energy in the near-horizon limit. This energy parameter plays an important role in our results and is defined in the main body of this paper in equation \eqref{Uenergy}. 

The FL bound was further sharpened in \cite{Montero:2021otb} by fixing the $\mathcal{O}(1)$ constant in the bound and by including thermal effects in \cite{Venken:2023hfa} in addition to other investigations in \cite{DallAgata:2021nnr,Lee:2021cor,Ban:2022jgm,Guidetti:2022xct,Montero:2022jrc,Mishra:2022fic,Mohseni:2022ftn,Cribiori:2023gcy,Dalianis:2023ewd,Abe:2023anf,Chrysostomou:2023jiv,Mohseni:2023ogd}, but we believe that there are still foundational questions that warrant further investigation. Foremost, the FL bound was derived in the probe limit where the mass and charge of the states are sufficiently small with respect to those of the black hole. Thus, the derivation of \cite{Montero:2019ekk} does not directly take into account higher orders in backreaction which, as we will argue, is of importance to understand the decay of charged black holes in de Sitter space. In comparison, to understand if it is possible to form a naked singularity by throwing a charged and/or spinning particle into a near-extremal Kerr-Newman black hole it is necessary to include backreaction effects. Historically, subleading backreaction effects in this process were ignored leading to ambiguities whether or not naked singularities can be formed this way \cite{Hubeny:1998ga}. However, subsequent analysis revealed that---when carefully taking into account second-order variations of thermodynamic quantities---naked singularities are not formed in this process \cite{Sorce:2017dst}. Similarly, we study in this article whether the subleading backreaction corrections to the decay of charged black holes in de Sitter space   modify the FL bound.

The analysis of \cite{Montero:2019ekk} considered the decay of Nariai black holes from a two-dimensional near-horizon perspective. By construction, this   neglects physical effects that are only visible away from a near-horizon limit    and it is therefore interesting to understand this decay directly from the four-dimensional theory. Indeed, we will see that there is a subtle rescaling of the four-dimensional energy in the near-horizon limit that needs to be taken into account.

The method we use to obtain a four-dimensional backreacted decay probability for charged black holes is the tunneling approach to Hawking radiation developed by \cite{Kraus:1994by,Kraus:1994fj,Parikh:1999mf}. Originally studied for uncharged radiation we adapt this method to study the decay of charged black holes by emission of charged radiation in asymptotically flat space (reviewing the earlier work \cite{Aalsma:2018qwy}) and de Sitter space. The benefit of this tunneling approach is that it directly  takes  into account all orders in backreaction of a classical spherical shell of radiation on the black hole geometry. For sufficiently heavy particles and ignoring the one-loop determinant, we find that the probability of a black hole to emit a charged particle is given by the universal formula 
\beq
\label{prob101}
P \sim e^{\Delta S_b} ~,
\eeq
both for black holes in flat space and in de Sitter space. Here $\Delta S_b$ is the entropy of   the  outer  black hole horizon after emission minus the entropy before emission. This probability has recently also appeared in the   study of  other gravitational tunneling processes \cite{Bah:2022uyz,Bintanja:2023vel}. Clearly,   \eqref{prob101} is only valid for decays satisfying~$\Delta S_b \leq 0$. The opposite regime $\Delta S_b>0$ corresponds to an instability known as charged superradiance \cite{Gibbons:1975kk} (for a review see~\cite{Brito:2015oca}). When considering superradiant decays, the probability distribution is modified to \cite{Aalsma:2018qwy}
\beq
\label{probsuper}
P_{\rm SR} \sim \frac{1-e^{\Delta S_b}}{(2-e^{-\Delta S_b})^2} ~.
\eeq
These expressions \eqref{prob101} and \eqref{probsuper} are valid for all black hole decays in flat and de Sitter space and we will show that, in the Nariai case, they give a generalization of the Schwinger pair production rate found by \cite{Montero:2019ekk} in the near-horizon limit. Using these decay probabilities it is now possible to study the subleading backreaction corrections to  charged black hole decay in de Sitter space. Doing so, we reach a striking conclusion: including higher orders in backreaction the onset of superradiance is modified and unsuppressed decays do not lead to a big crunch for every observer. Moreover, in the near-horizon limit superradiance is completely decoupled for real near-horizon mass parameter $m_s$.   This reveals that the unsuppressed decay channels we find using the tunneling approach violate the FL bound, but have an end state as a standard Reissner-Nordstr\"{o}m-de Sitter black hole.

In addition, we show that $\Delta S_b \leq -1$  reduces to the FL bound in the probe limit $H/M_p \ll 1$,  and we compute the first subleading correction to the FL bound. When the bound $\Delta S_b \leq -1$ is violated, decays are unsuppressed and the electric field of the black hole is discharged rapidly. In the probe limit and in the effective near-horizon  Nariai geometry, \cite{Montero:2019ekk} argued that this causes a near-horizon observer to end up in a big crunch singularity, effectively because the black hole horizon passes them in this process.\footnote{Charged Nariai black holes have   an event horizon  radius that is smaller than  that of  an uncharged Nariai black hole if the   de Sitter curvature radius is kept fixed.}  Taking higher-order backreaction corrections into account we identify unsuppressed decay channels via the tunneling approach for which this is not the case, suggesting the possibility that the decay process does not result in a big crunch singularity, although a definitive answer relies on whether this decay channel is dominant or subdominant with respect to the multi-particle decays found in \cite{Montero:2019ekk}.

Because  such decays  result in a standard Reissner-Nordstr\"{o}m-de Sitter geometry, even though they could violate $\Delta S_b \leq -1$, there is a   possibility that an observer remains outside of the black hole horizon and does not witness a big crunch. We will not explicitly analyze the possible formation of a big crunch singularity in this article, as was done in \cite{Montero:2019ekk},  and leave its analysis   in four dimensions away from the near-horizon and probe limit of unsuppressed decay to future work. 

The layout of this paper is as follows. In Sec. \ref{sec:BHsolutions} we introduce our notation and conventions and describe the charged black hole solutions of interest. Then, in Sec.~\ref{sec:Tunneling} we use the tunneling method to derive the probability of a charged black hole in flat and de Sitter space to emit a massive, charged state. We compare and contrast the superradiant regime for both black hole solutions. In Sec. \ref{sec:Bounds} we study how backreaction corrections modify the conclusion that all unsuppressed decays from Nariai black holes lead to a crunching geometry for every observer. Further, we take   extremal near-horizon  limits of the decay probabilities in Sec. \ref{sec:NHlimits} demonstrating that we recover the correct expressions for Schwinger pair production in the different   limits.  We also show how imposing $\Delta S_b \leq -1$ in the probe limit leads to the exact FL bound including the correct numerical factors. Finally, we conclude in Sec. \ref{sec:Conclusion}.

\section{Electrically Charged Black Hole Solutions} \label{sec:BHsolutions}
We start with a brief review of Reissner-Nordstr\"{o}m black hole solutions in asymptotically flat space and  de Sitter space in four spacetime dimensions.

\subsection{Reissner-Nordstr\"{o}m Black Holes} \label{sec:flatRN}
The Einstein-Maxwell action with zero cosmological constant is
\beq
\int\rmd^4x\sqrt{-g}\left(\frac1{16\pi G_4}R - \frac14 F_{ab}F^{ab}\right) ~.
\eeq
The equations of motion associated with this action are given by the Einstein field equation and sourceless Maxwell   equation
\beq
\bal
R_{ab} - \frac12g_{ab}R &= 8\pi G_4 T_{ab}~, \\
\nabla_aF^{ab} &=0 ~,
\eal
\eeq
where the electromagnetic stress tensor takes the form 
\beq \label{eq:EMstress}
T_{ab} = F_{ac}F_b^{\,\,\,c} - \frac14g_{ab}F_{cd}F^{cd} ~.
\eeq
The charged,   static, spherically symmetric solution to the equations of motion describes the asymptotically flat Reissner-Nordstr\"{o}m (RN) black hole. The   RN line element and gauge field are    
\beq
\bal
\label{RNmetric}
\rmd s^2 &= -f(r)\rmd t^2 + f(r)^{-1}\rmd r^2 + r^2\rmd \Omega_2^2 ~,\\
f(r)&= 1 -\frac{2G_4M}{r} + \frac{G_4Q^2}{4\pi r^2} ~, \\
A&=  \left(-\frac{Q}{4\pi r}+\lambda\right)\rmd t ~.
\eal
\eeq
Here ($M,Q$) are the mass and electric charge of the black hole,  and $\lambda$ is a gauge parameter. Moreover, the field strength is given by $F=\rmd A$.

The roots of $f(r)$ determine the location of the inner (Cauchy) horizon and outer (event) horizon, $r=r_-$ and $r=r_+$, respectively, which in terms of the mass and charge are given by
\beq
r_\pm = G_4M \pm \sqrt{G_4^2M^2-\frac{G_4Q^2}{4\pi}} ~.
\eeq
All regular black holes obey $r_+\geq r_-$, or, equivalently, the extremality bound (in ADM units)
\beq \label{extremalityRN}
M\geq \frac{Q}{2\sqrt{\pi G_4}} ~.
\eeq
A Reissner-Nordstr\"{o}m black hole is called `extremal' if this bound is saturated. For future convenience, we define the mass and charge parameters
\beq \label{eq:masschargepara}
{\cal M} = 2G_4M ~, \quad {\cal Q} = \sqrt{\frac{G_4}{4\pi}} Q ~,
\eeq
such that the extremality bound \eqref{extremalityRN} turns into ${\cal M}\geq 2{\cal Q}$. Later, when we refer to the mass   and charge of states emitted by the black hole we indicate that with lowercase $(m,q)$ in ADM units and $(\mathfrak{m},\mathfrak{q})$ for the units \eqref{eq:masschargepara}. For example the WGC in these units is:
\beq
m \leq \frac{q}{2\sqrt{\pi G_4}} \quad \text{or} \quad \mathfrak{m} \leq 2\mathfrak{q} ~. 
\eeq
Further, the Hawking temperature and Bekenstein-Hawking entropy associated to the event horizon at $r=r_+$ are 
\beq
T =  \frac{r_+-r_-}{4\pi r_+^2} \,, \qquad S = \frac{4\pi r_+^2}{4G_4}~.
\eeq
Finally, it is well-known that the near-horizon geometry of extremal Reissner-Norstr\"{o}m black holes is given by AdS$_2\times S^2$. This can be see by defining near-horizon coordinates $(\tau,\zeta)$ through
\beq
r = r_+ + \epsilon \frac{r_+^2}{\zeta} ~, \quad t = \frac{\tau}{\epsilon} ~.
\eeq
First taking the extremal limit in the metric and subsequently expressing it in terms of $(\tau,\zeta)$,    the RN line element \eqref{RNmetric} becomes in the near-horizon limit $\epsilon\to0$
\beq
\label{ads2nearhorizon}
\rmd s^2 = \frac{r_+^2}{\zeta^2}\left(-\rmd\tau^2+\rmd\zeta^2\right) + r_+^2\rmd\Omega_2^2 ~.
\eeq
We recognize this as the metric of AdS$_2\times S^2$ in Poincaré coordinates with an equal AdS$_2$ and $S^2$ radius $\ell_2=r_+$.

\subsection{Reissner-Nordstr\"{o}m-de Sitter Black Holes}
We now turn our attention to the considerably more rich  charged black hole solutions in asymptotically de Sitter space (see e.g. \cite{Romans:1991nq,Mann:1995vb,Morvan:2022aon}). The Einstein-Maxwell action with positive cosmological constant $\Lambda_4 = 3/
\ell_4^2$ is given by
\beq
\int\rmd^4x\sqrt{-g}\left[\frac1{16\pi G_4}\left(R-\frac6{\ell_4^2}\right) - \frac14 F_{ab}F^{ab}\right] ~.
\eeq
Here $\ell_4$ is the four-dimensional de Sitter radius. The equations of motion take a similar form as before
\beq
\bal
R_{ab} - \frac12g_{ab}R + \frac3{\ell_4^2}g_{ab} &= 8\pi G_4T_{ab} & ~,\\
\nabla_aF^{ab} &=0 ~.
\eal
\eeq
The stress tensor is again given by \eqref{eq:EMstress}. The field equations admit a charged, static, spherically symmetric solution with de Sitter asymptotics, which we refer to as the Reissner-Nordstr\"{o}m-de Sitter (RNdS) solution,
\beq \label{eq:metricdef}
\bal
\rmd s^2 &= -f(r)\rmd t^2 + f(r)^{-1}\rmd r^2 + r^2\rmd \Omega_2^2 ~,\\
f(r)&= 1  - \frac{r^2}{\ell_4^2} -\frac{2G_4M}{r} + \frac{G_4Q^2}{4\pi r^2} ~, \\
A&= \left(-\frac{Q}{4\pi r}+\lambda\right)\rmd t ~.
\eal
\eeq
As before, $(M,Q)$ are the mass and charge of the black hole.

\begin{figure}[t]
\centering
\includegraphics[scale=.9]{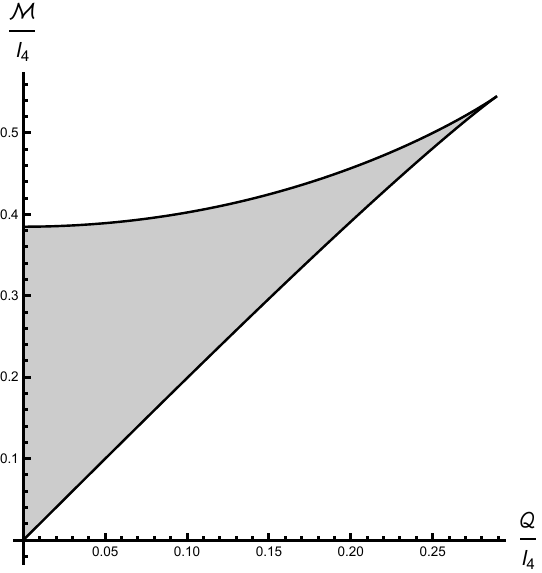}
\caption{Shark fin-shaped   diagram for Reissner-Nordstr\"{o}m-de Sitter black holes. The shaded region corresponds to black hole solutions, and the white region to solutions with naked singularities.   The lower boundary of the shark fin corresponds to cold black holes $(r_a=r_b)$ and the upper curved boundary  to Nariai black holes $(r_b=r_c)$. The two lines meet at the ultracold point $r_a=r_b=r_c$.}
\label{fig:sharkfin}
\end{figure}

The parameter space of this solution is much more intricate than for charged black holes in flat space. We can see this by writing the blackening factor in terms of the roots of $f(r)$ as  
\beq \label{eq:ChargeddS}
f(r) = -\frac{(r-r_c)(r-r_b)(r-r_a)(r+r_a+r_b+r_c)}{r^2\ell_4^2} ~.
\eeq
We notice that $f(r)$, being a quartic polynomial, has 4 roots given by
\beq
r = r_i ~, \quad r = -(r_a+r_b+r_c) ~.
\eeq
To identify the locations of the different horizons, we can impose $r_c\geq r_b\geq r_a \geq 0 $. This implies that the root $r = -(r_a+r_b+r_c)$ is unphysical and we can identify $r_a$ as the location of the inner horizon of the black hole, $r_b$ as the outer horizon of the black hole and $r_c$ as the cosmological horizon. 
The RNdS black hole is fully characterized by the three independent parameters $M, Q$ and $\ell_4$, which can be expressed as a function of the three horizon radii. For instance, we can relate the de Sitter radius to the three horizon radii as follows
\beq \label{eq:dSradius}
\ell_4^2 = r_a^2+r_b^2+r_c^2 + r_ar_b+r_br_c+r_ar_c ~.
\eeq
The space of black hole  solutions can   be determined by considering its boundaries: the regions where extremal solutions exist that lead to a double root of the blackening factor $f(r)$. These can be found by considering the discriminant of $f(r)$
\beq
\text{Disc}\left[f(r)\ell_4^2r^2\right] = (r_a-r_b)^2(r_b-r_c)^2(r_a-r_c)^2(2r_a+r_b+r_c)^2(r_a+2r_b+r_c)^2(r_a+r_b+2r_c)^2 ~.
\eeq
The discriminant is zero if and only if $f(r)$ has a double root. Following the terminology in \cite{Romans:1991nq} we call the  three extremal solutions
\beq
\bal
\text{Cold Black Hole:} \quad r_a &= r_b  ~, \\
\text{Charged Nariai Black Hole:} \quad r_b &= r_c  ~, \\
\text{Ultracold Black Hole:} \quad r_a &= r_b =  r_c ~.
\eal
\eeq
The cold black hole is the de Sitter analogue of the extremal RN black hole in asymptotically flat space. The (charged) Nariai black hole is the largest possible black hole that fits inside the cosmological horizon. The ultracold black hole corresponds to the solution   for which    the cold black hole becomes as large as the cosmological horizon. Together, these three solutions determine the boundaries of the phase diagram of regular charged black holes in de Sitter space. This phase diagram is commonly referred to as the `shark fin', see Figure~\ref{fig:sharkfin}.

Using   relation \eqref{eq:dSradius}, the mass and charge can be expressed purely in terms of two horizon radii and  $\ell_4$ 
\beq
\bal \label{eq:dSmasscharge}
M &= \frac{1}{2G_4}\frac{r_b^2(r_b^2-\ell_4^2)-r_c^2(r_c^2-\ell_4^2)}{(r_c-r_b)\ell_4^2} ~,\\
Q^2 &= \frac{4\pi}{G_4}\frac{r_c(r_c^2-\ell_4^2)-r_b(r_b^2-\ell_4^2)}{(r_c^{-1}-r_b^{-1})\ell^2_4} ~. 
\eal
\eeq
We again define the mass and charge parameters $({\cal M},{\cal Q})$ as in \eqref{eq:masschargepara}. The relation \eqref{eq:dSmasscharge} between the mass, charge and horizon radii can be inverted to yield a complicated, but analytic   expression for the horizon radii (see Appendix A in \cite{Morvan:2022aon})
\beq \label{eq:ExplicitRadii}
r_{b,c} = r_U \left(x\mp \sqrt{3-x^2-\frac2x\frac{{\cal M}}{{\cal M}_U}}\right) ~,
\eeq
where we introduced
\beq
\bal
x &= \sqrt{1+\sqrt{1-\rho^2}\cos(2\eta)} ~, \\
\eta &= \frac13\arccos\left(\sqrt{\frac{{\cal M}_C^2-{\cal M}^2}{{\cal M}_C^2-{\cal M}_N^2}}\right) ~, \\
\rho &=q/q_U~. 
\eal
\eeq
Here $r_U = \ell_4/\sqrt{6}$ denotes the radius of the ultracold black hole, and ${\cal M}_U=\ell_4\left(\frac23\right)^{3/2}$ and ${\cal Q}_U = 
\ell_4 / (2 
\sqrt{3})$ are the mass and charge of the ultracold black hole, respectively. Further,
\beq \label{eq:NariaiColdMass}
{\cal M}_{N,C} = \frac23 r_U \sqrt{1+3\rho^2\pm(1-\rho^2)^{3/2}} 
\eeq
corresponds to the mass of the Nariai and cold black hole. For the special case of Nariai and cold black holes the expression for the horizon radii simplifies to
\begin{equation}
    r_{N,C} = r_U \sqrt{1\pm \sqrt{1- \rho^2}}\,. 
\end{equation}
Here $r_N$ (plus sign on the right-hand side) is the radius of a charged  Nariai black hole and $r_C$ (minus sign on the right-hand side) the radius of a cold black hole.

The presence of different kinds of extremal RNdS black holes  yields a rich structure of   extremal limits and associated near-horizon geometries that we will describe in Sec.~\ref{sec:NHlimits}. The thermodynamics of the near-extremal limits has recently been studied in detail in~\cite{Castro:2022cuo}.

\section{Black Hole Decay as a Tunneling Process} \label{sec:Tunneling}
To derive the probability rate of charged particles tunneling through the black hole and/or cosmological horizon we follow the approach of Parikh and Wilczek \cite{Parikh:1999mf}. Our starting point is the (quantum-mechanical) amplitude for a black hole emitting a particle in a classically forbidden transition. Focussing on the dominant spherically symmetric (s-wave) channel and assuming validity of the WKB approximation, this amplitude is proportional to the exponent of the classical on-shell action $I(m,q)$ of a spherical shell of mass $m$ and charge $q$
\beq
{\cal A}_{m,q} \propto e^{i I(m,q)} ~.
\eeq
This expression does not include the  one-loop determinant  multiplying the exponent, that can be determined by perturbing around the saddle point. We will ignore the one-loop determinant in this article, since we think the exponential behavior is the most relevant for the Festina Lente bound.

From this amplitude, the probability can be determined making it clear that only the imaginary part of the action contributes
\beq
P_{m,q} = |{\cal A}_{m,q}|^2 \propto e^{-2\text{Im}(I(m,q))} ~.
\eeq
Only classically forbidden trajectories give a non-zero amplitude, allowing an interpretation of this transition as a tunneling process. We will now determine the on-shell action in a generic spherically symmetric background (for more details on the derivation   see \cite{Aalsma:2018qwy}), and later we specialize to the RN(dS) geometry.

\subsection{On-shell Action}
The   on-shell action of classical matter can always be expressed in the Hamiltonian formalism as
\beq
I = \int \rmd \mathfrak{t} \,(P \dot X - {\cal H}) ~.
\eeq
Here $\mathfrak{t}$ is an appropriate time coordinate used to determine the evolution of the shell, $(X,P)$ are the canonical position and momentum, respectively, and ${\cal H}$ is the Hamiltonian density. The dot denotes a derivative with respect to $\mathfrak{t}$. Because we focus on a spherically symmetric shell, its trajectory is determined by $X =  r( \mathfrak{t})$. Only the first term in the action contributes to the tunneling probability, since   the Hamiltonian is real. Then, using Hamilton's equation
\beq
\dot r = \frac{\partial{\cal H}}{\partial P} ~,
\eeq
the imaginary part of the action is given by
\beq
\text{Im}(I) = \text{Im}\left(\int \rmd{\cal H}\int\frac{\rmd r}{\dot r}\right) ~.
\eeq
Clearly, an imaginary contribution could arise from poles in $\dot r^{-1}$ which can be determined using the equations of motion. Since we are interested in matter tunneling through a horizon, it is convenient to express the equations of motion in coordinates that are regular there. In particular, starting with a generic spherically symmetric metric of the form
\beq
\rmd s^2 = -f(r)\rmd t^2 + f(r)^{-1}\rmd r^2 + r^2\rmd\Omega_2 ^2 ~,
\eeq
we introduce Painlevé-Gullstrand coordinates. They are defined by a   transformation of the time coordinate 
\beq
\mathfrak{t}=  t \pm \int\rmd r\frac{\sqrt{1-f(r)}}{f(r)} ~,
\eeq
which leads to the metric
\beq 
\rmd s^2 = -f(r)\rmd\mathfrak{t}^2 \pm 2\sqrt{1-f(r)}\rmd\mathfrak{t} \rmd r +\rmd r^2+ r^2\rmd\Omega_2^2 ~.
\eeq
In Painlevé-Gullstrand coordinates the metric is regular at $f(r)=0$.  Assuming $\mathfrak{t}$ increases towards the future, the upper sign is appropriate to describe outgoing matter moving towards larger $r$. Conversely, the lower sign is appropriate to describe incoming shells moving to smaller $r$. Because we are interested in radiation emitted by black holes, we focus   on outgoing shells and therefore choose the upper sign.

\begin{figure}[t]
\centering
\includegraphics[scale=.65]{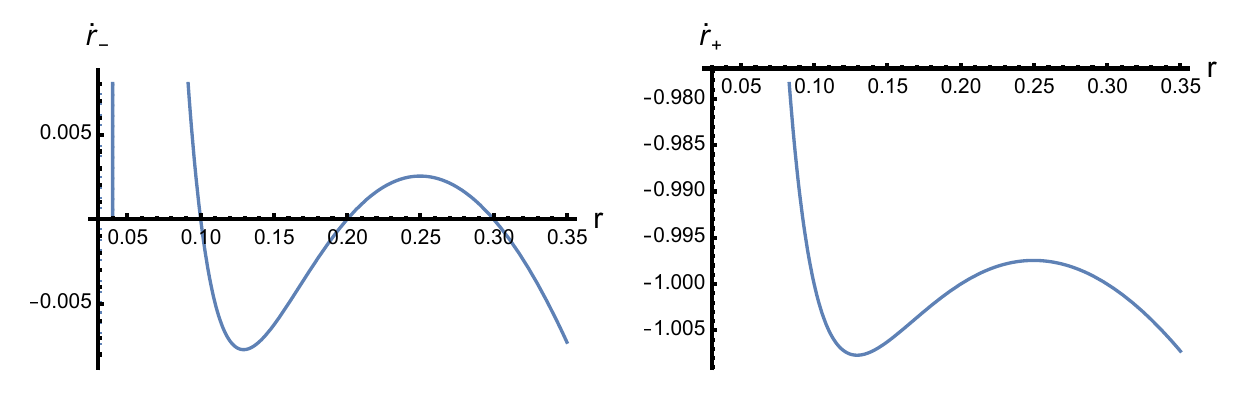}
\caption{The two trajectories $\dot r_\pm$ in RNdS for $r_a=0.1,r_b=0.2,r_c=0.3$, $E=1$, and $\ell_4=1$. Positive $\dot r$ corresponds to an outgoing trajectory and negative $\dot r$ to an incoming trajectory. The fact that $\dot r_-=0$ at the three horizon radii indicates that a trajectory travelling through a horizon is classically forbidden and therefore describes a tunneling process. $\dot r_-$ describes an outgoing shell in the region $r_b<r<r_c$ that has tunneled through the outer black horizon. Both trajectories $\dot r_\pm$ vanish at $f(r)=E^2$. }
\label{fig:trajectories}
\end{figure}

We now consider a timelike trajectory of the outgoing matter, parametrized by $x^a(\mathfrak{t})$. Denoting the proper time along this trajectory as $s$, we can define the four-velocity with unit normalization as $u^a=\frac{dx^a}{ds}$. Using the explicit form of the metric, the normalization condition $u_au^a=-1$ can be written as
\beq \label{eq:fourvelocitynorm}
f(r) - 2 \sqrt{1-f(r)}\dot r -\dot r^2 = \dot s^2 ~.
\eeq
One of the advantages of Painlevé-Gullstrand coordinates is that the metric has  a manifest timelike Killing vector given by $K = \partial_\mathfrak{t} $. Using this, we can identify a conserved quantity $E$ along the trajectory
\beq
E = K_a\frac{dx^a}{ds} \quad \to \quad E\dot s = -f(r) + \sqrt{1-f(r)}\dot r ~.
\eeq
Plugging this into \eqref{eq:fourvelocitynorm} we can solve for $\dot r$ to find two possible trajectories
\beq  \label{eq:rdotsolution}
\dot r_\pm = \frac{\sqrt{1-f(r)}f(r) - E^2\left(\sqrt{1-f(r)} \pm \sqrt{1-\frac{f(r)}{E^2}}\right)}{1+E^2-f(r)} ~.
\eeq
Their physical interpretation can be best seen by plotting them for a specific solution. For example, for a Reissner-Nordstr\"{o}m-Sitter black hole the blackening factor is given by \eqref{eq:ChargeddS} and the two trajectories are displayed in Figure \ref{fig:trajectories}.
In terms of $f(r)$, the solution $\dot r_-$ of \eqref{eq:rdotsolution} has two roots given by $f(r)=0$ and $f(r)=E^2$ whereas $\dot r_+$ only has one root given by $f(r)=E^2$. Only $\dot r_-$ corresponds to a trajectory that tunnels through the  outer  black hole horizon, so we focus on that solution.  

As we will see next, the roots corresponding to the location of the horizon ($f(r)=0$) will lead to a nonzero contribution to the imaginary part of the action. This contribution can be computed by specifying the specific form of $f(r)$.

\subsection{Tunneling Probability}
We will now consider the tunneling probability for the emission of charged shells from charged black holes in flat and de Sitter space.

\subsubsection*{Reissner-Nordstr\"{o}m Black Holes}
First, we review the computation of the tunneling probability for charged shells emitted by Reissner-Nordstr\"{o}m black holes, which was performed in \cite{Aalsma:2018qwy}. We consider a RN black hole with initial mass and charge $(M,Q)$ that decays to a RN black hole with  smaller final mass and charge $(M-m,Q-q)$.  We thus focus on shells with positive energy and positive charge. The black hole decay is then mediated by a tunneling process in which an outgoing shell of spherically symmetric matter starts inside but close to the outer horizon $r_+$, tunnels through it, and emerges just outside. Afterwards, the shell follows a classical trajectory. 
The imaginary part of the action is thus given by
\beq
\text{Im}(I) = \text{Im}\left(\int_{M}^{M-m}\rmd{\cal H}\int_{r_+-\epsilon}^{r_++\epsilon}\frac{\rmd r}{\dot r}\right) ~,
\eeq
with $\epsilon$ small and positive. Here we used that the Hamiltonian is given by the mass of the black hole. This integral can be computed by using  \eqref{eq:rdotsolution} (the --  solution) and inserting the  blackening factor for RN black holes, $f(r) =  (r-r_+)(r-r_-) /r^2 $.

To evaluate the integral  we need to deform the contour as $m\to m-i\epsilon$ (for $\epsilon > 0$) for future propagation of positive energy modes. In \cite{Aalsma:2018qwy} it was argued this is the correct contour deformation. In short, when deriving the propagator on a curved background this deformation yields the Feynman propagator. With this prescription, the first integral evaluates to $-i\pi$ times the sum of the residues at the poles 
\beq
 \text{Im}\left(\int_{r_+-\epsilon}^{r_++\epsilon}\frac{\rmd r}{\dot r}\right) = - i\pi\text{Res}\left(\frac1{\dot r}\right)_{r=r_b} ~.
\eeq
There exists a pole at the black hole event horizon, whose residue is given by 
\beq
\bal
\text{Res}\left(\frac1{\dot r}\right)_{r=r_b} &= \frac{2r_+^2}{r_+-r_-} ~.
\eal
\eeq
Finally, we need to know the change of the Hamiltonian while keeping the charge fixed. Using the definition of mass and charge in terms of horizon radii we find
\beq
\rmd {\cal H} =  \frac{r_+-r_-}{2G_4r_+}\rmd r_+
\eeq
Using this relation, we can easily perform the last integral to find the final expression 
\beq
\text{Im}(I) = -\frac{\pi}{2G_4}\Delta r_+^2 ~.
\eeq
Here $\Delta r_b $ is defined as the horizon radius after emission minus the radius before emission. We recognize this as minus one half times the total entropy difference of the black hole event horizon. Thus, we find that
\beq \label{eq:RNtunneling}
P_{m,q} = e^{-2\text{Im}(I(m,q))} = e^{\Delta S_b(m,q)} ~.
\eeq
Here $\Delta S_b$ equals the entropy after emission minus the entropy before emission.

\subsubsection*{Reissner-Nordstr\"{o}m-de Sitter Black Holes}
Next, we will determine the tunneling probability for the emission of outgoing charged shells from  Reissner-Nordstr\"{o}m-de Sitter black holes. As before, we consider spherical matter with positive mass $ m$ and charge $ q$ that starts out inside the outer horizon at $r=r_b-\epsilon$, tunnels through the outer black hole horizon and possibly moves classically through the cosmological horizon.   Because this latter part of the evolution is classical it does not contribute to the imaginary part of the action. In principle, there is also the possibility of a process in which an incoming positive energy shell tunnels from outside the cosmological horizon and is absorbed by the black hole. Such a process would be governed by the change in entropy of the cosmological horizon. However, fixing the final state of the black hole to have lower mass, such a process would need to be accompanied by emission of an outgoing shell by the black hole. Such multi-particle decays are subdominant---at least away from the superradiant regime defined later on---and we therefore opt not to study this process. 

The imaginary part of the action is given by
\beq
\text{Im}(I) = \text{Im}\left(\int_{M}^{M-m}\rmd{\cal H}\int_{r_b-\epsilon}^{r_b+\epsilon}\frac{\rmd r}{\dot r}\right) ~,
\eeq
Again, this integral can be computed by using   \eqref{eq:rdotsolution} (the --  solution), but now by inserting the blackening factor \eqref{eq:ChargeddS} of the RNdS black hole.
Using the same contour prescription, we find that
\beq
 \text{Im}\left(\int_{r_b-\epsilon}^{r_b+\epsilon}\frac{\rmd r}{\dot r}\right) = - i\pi\,\text{Res}\left(\frac1{\dot r}\right)_{r=r_b} ~.
\eeq
The residue  at the horizon  is given by
\beq
\text{Res}\left(\frac1{\dot r}\right)_{r=r_b} =  \frac{2\ell_4^2r_b^2}{(r_b-r_a)(r_c-r_b)(r_a+2r_b+r_c)} ~.
\eeq
Further, the change in Hamiltonian at fixed charge is  
\beq
\rmd {\cal H} =  \frac{(r_a-r_b)(r_b-r_c)(r_a+2r_b+r_c)}{2G_4\ell^2 r_b}\rmd r_b ~.
\eeq
The last integral now yields
\beq
\text{Im}(I) = -\frac{\pi}{2G_4}\Delta r_b^2\,. 
\eeq
We recognize this as minus one half times the entropy difference of the outer black hole horizon. Thus, the probability rate is
\beq \label{eq:dStunneling}
P_{m,q} = e^{-2\text{Im}(I(m,q))} = e^{\Delta S_b(m,q) } ~,
\eeq
where $\Delta S_{b}$ equals the entropy after emission minus the entropy before emission. Here we only considered the probability of an outgoing shell with positive mass and charge tunneling through the black hole horizon.   Instead, one could study an incoming shell with positive mass and charge tunneling through the cosmological horizon. The probability rate for such a tunneling process can be computed in a similar fashion as above and  is equal to $P\sim \exp(\Delta S_c).$   Below we will just focus on decay processes that are described by an outgoing shell of positive mass and charge (i.e. with positive $\dot r$). This is sufficient for our purposes and we show that this reproduces existing results in the literature, such as the Schwinger pair production rate in the different extremal near-horizon limits. Still, it would be interesting to understand  the role of decays described by incoming shells in more detail,  which we leave for future study (see also \cite{Medved:2002zj}).

\subsection{Superradiant Decays}
\label{sec:superradiant}
Clearly, we can only interpret the derived expressions as (normalized) decay probabilities when the entropy difference is negative. Instead, whenever the entropy difference is positive the expression for the probability is not bounded by one and needs to be reinterpreted. As argued in \cite{Aalsma:2018qwy}, for a Reissner-Nordström black hole the cross-over to the regime $\Delta S_b > 0$ can be explained in terms of charged superradiance: a rapid decay process in which spontaneous production of (light) superradiant modes is entropically favorable. This leads to a fast discharge of the black hole. The probability distribution for superradiant decays is modified in the following way \cite{Aalsma:2018qwy}
\beq 
\Delta S_b > 0: \quad P_{\rm SR} \sim \frac{1-e^{\Delta S_b}}{(2-e^{-\Delta S_b})^2} ~.
\eeq
  In de Sitter space  we also define the superradiant regime   as the region where $\Delta S_b > 0$. This agrees with the thermodynamic definition of superradiance as $dM >0$ \cite{Brito:2015oca} because of  the first law for the outer black hole horizon at fixed charge, i.e. $dM=T_bdS_b$. We stress that this probability is valid for the emission of a single superradiant mode, assuming an underlying Bose-Einstein distribution that ignores interactions between particles. However, one might argue that in the superradiant regime it will become favorable to emit a large number of particles such that interactions can no longer be ignored.  However, that is mostly irrelevant for our purposes in this paper, which is mostly concerned with the non-superradiant ($\Delta S_b < 0$) regime. Clearly, the single-particle assumption is valid in the regime $\Delta S_b \ll -1$ where it only ignores small corrections, but here we will entertain the possibility that the validity of \eqref{eq:dStunneling} extends to the regime $-1<\Delta S_b <0$.  Although it would be interesting to study multi-particle decay more in detail, in this article it suffices to focus on decays mediated by a single shell which are the dominant contribution away from the superradiant regime. As we will see later on, in the near-horizon extremal limits that are the main interest of this article the superradiant regime completely decouples. The fact that we reproduce existing results for Schwinger pair production in the extremal limit justifies our assumption that the dominant contribution to the decay is mediated by a single shell.

We will now give explicit expressions for the superradiant regime for different extremal black holes highlighting their similarities and stressing their differences.

\subsubsection*{Charged Superradiance in Flat Space}
To derive explicit expressions for the entropy difference, we find it useful to work with the following mass and charge parameters for emitted states:
\beq
\mathfrak{m} = 2G_4m ~, \qquad \mathfrak{q} = \sqrt{\frac{G_4}{4\pi}} q ~.
\eeq
If we start with an extremal RN black hole, which has $({\cal M},{\cal Q}) = (2{\cal Q},{\cal Q})$ in our conventions, and consider emission of a shell with mass and charge parameter $(\mathfrak m, \mathfrak q)$, the parameter region where superradiance occurs is given by
\beq
\label{superradiant1}
\Delta S_b > 0: \qquad \mathfrak{m} < 2\mathfrak{q} - \frac{\mathfrak{q}^2}{{\cal Q}} ~.
\eeq
As noted in \cite{Aalsma:2018qwy}, the fact that we take into account higher orders in backreaction causes the superradiance condition and the WGC condition to differ. For the extremal charged black hole to decay without forming a naked singularity the shell needs to satisfy $\mathfrak{m}\leq 2\mathfrak{q}$, which corresponds to the WGC. Thus, in the probe limit $\mathfrak{q}/{\cal Q} \ll 1$ the second term in the superradiant condition \eqref{superradiant1} is subdominant and the WGC becomes equivalent to the superradiance condition. However, by taking into account second-order backreaction effects the tunneling computation opens up a regime of non-superradiant decays that nonetheless satisfy the WGC: $2\mathfrak{q} - \mathfrak{q}^2/{\cal Q} < \mathfrak{m} \leq 2\mathfrak{q}$.

This becomes relevant when we consider the expression for the decay rate in terms of the near-horizon region described by AdS$_2\times S^2$. For an observer in the near-horizon throat region, the time coordinate and therefore the definition of energy is rescaled compared to the time and energy of   an asymptotic observer. It therefore becomes necessary to define a new energy parameter as follows.  The stress-energy tensor $S_{ab}$ of the shell should satisfy the Israel junction conditions and is   given by \cite{Poisson:2009pwt}
\beq
S_{ab} = \frac{1}{8\pi G_4}\left([K_{ab}]-[K]h_{ab}\right) ~.
\eeq
Here the brackets $[A]=A_+-A_-$ indicate the difference between the quantity outside and inside the shell. $K_{ab}$ is the extrinsic curvatature with trace $K$ and $h_{ab}$ is the induced metric on the shell. The shell separates two geometries with different blackening factors: $f_+(r)$ outside the shell (Reissner-Nordström before emission) and $f_-(r)$ inside the shell (Reissner-Nordström after emission).   In other words, if we fix the ADM charge (as in \cite{Parikh:1999mf}), $f_+(r)$ is the blackening factor with mass $M$ and charge $Q$, whereas $f_-(r)$ is the blackening factor with mass $M-m$ and charge $Q-q.$   From the stress-energy tensor we can define  a tension ${\cal T}$ and pressure ${\cal P}$ as follows (see e.g \cite{Danielsson:2017riq})
\beq
\bal \label{eq:tensiondef}
{\cal T}(r) &:= S^t_{\,\,\,t} = \frac1{4\pi G_4 r}\left[\sqrt{f_-(r)}-\sqrt{f_+(r)}\right]~, \\
{\cal P}(r) &:= S^\theta_{\,\,\,\theta} = \frac1{16\pi G_4 r}\left[2\left(\sqrt{f_-(r)}-\sqrt{f_+(r)}\right)+r\left(\frac{f_-'(r)}{\sqrt{f_-(r)}}-\frac{f_+'(r)}{\sqrt{f_+(r)}}\right)\right] ~.
\eal
\eeq
These expressions have to be evaluated at  the location of the shell and the prime indicates a derivative with respect to~$r$. The total energy of the spherical shell at radius $r$ is now defined as
\begin{equation}
\label{Uenergy}
m_s(r) := 4\pi r^2 {\cal T}(r)\,.
\end{equation} If the initial state before emission is an extremal RN black hole, we   find that asymptotically this energy is given by
\beq
\lim_{r\to\infty} m_s(r) = \frac{\mathfrak{m}}{2G_4} = m ~.
\eeq
This is the standard definition of energy with respect to an asymptotic observer. Instead, at the horizon we obtain for the shell energy emitted by an extremal black hole: 
\beq \label{eq:NHenergyRN}
\lim_{r\to r_+} m_s(r) = \frac{\sqrt{{\cal Q}(\mathfrak{m}-2\mathfrak{q})+\mathfrak{q}^2}}{G_4} = \frac{\sqrt{f_-(r_+)}}{4\pi G_2\ell_2}~.
\eeq
Here $f_-(r_+)$ denotes the blackening factor inside the   shell  evaluated at  a root of the blackening factor $f_+(r)$ outside the shell, namely the outer horizon radius $r_+$. In the last equality we expressed the energy in terms of the two-dimensional Newton constant $G_2=G_4(4\pi\ell_2^2)^{-1}$   and the AdS$_2$ radius $\ell_2=r_+$, which is equal to the $S^2$ radius (see \eqref{ads2nearhorizon}). This energy scales with the AdS$_2$ radius as a Kaluza-Klein mode and the appearence of the blackening factor captures the rescaling of the energy in a manner appropriate for a near-horizon observer.

Thus, evaluated at a general radial location, the energy parameter $m_s$ interpolates between the standard definition of asymptotic energy at infinity and the horizon energy. The observer with respect to which we define the energy determines which energy parameter is relevant. For example, for the (four-dimensional) WGC one defines states asymptotically, identifying $m$ as the relevant energy. On the other hand, we can also consider decays of the black hole from an AdS$_2\times S^2$ near-horizon perspective. In that case, it is more natural to use \eqref{eq:NHenergyRN} as the definition of energy. Indeed, expressing $\Delta S_b$ in terms of $m_s(r_+)$,\footnote{For ease of notation, unless indicated otherwise, $m_s$ will always refer to the  horizon energy.} and taking the limit $\sqrt{G_4}/\ell_2\ll 1$ the entropy difference becomes~\cite{Aalsma:2018qwy}
\beq
\Delta S_b= -2\pi \ell_2^2\left(qE - \sqrt{(qE)^2 - (m_s/\ell_2)^2}\right) + {\cal O}(\sqrt{G_4}/\ell_2) ~,
\eeq
where we introduced the electric field in the near-horizon limit
\beq
E = \frac{Q}{4 \pi r_+^2}  = \frac1{\sqrt{4\pi G_4}\ell_2} ~.
\eeq
In the second equality we inserted $Q = {\cal Q} \sqrt{4\pi / G_4}$, $r_+ = \ell_2$ and ${\cal Q }= \ell_2$.
This leads to a decay rate
\beq \label{eq:AdS2Schwinger}
\Gamma \sim \exp\left(-2\pi \ell_2^2\left(qE - \sqrt{(qE)^2 - (m_s/\ell_2)^2}\right)\right) ~,
\eeq
that coincides with Schwinger pair production in AdS$_2$ \cite{Pioline:2005pf,Kim:2008xv}. 

One interesting aspect of the  horizon energy is that the superradiant regime is decoupled for positive energies $m_s>0$. Decays for which $\Delta S_b = 0$ correspond to ${\cal T}=0$. Still, because we took into account backreaction there is a regime of non-superradiant emission that satisfies the WGC and can be described by Schwinger pair production in AdS$_2$. We stress that although for positive $m_s$ superradiance is decoupled, this does not mean that such decays are forbidden. Instead, they have a well-defined description in the asymptotic region where we express energy in terms $m$, but lead to imaginary $m_s$ when we evaluate this energy in the near-horizon region. In the probe limit where $\mathfrak{q}/{\cal Q}\ll 1$---which coincides with $  q\sqrt{G_4}/\ell_2 \ll 1$, since $\mathfrak{q}\sim \sqrt{G_4}q$ and ${\cal Q}= r_b = \ell_2$---the distinction between superradiance and the WGC disappears so that all decays are superradiant.

\subsubsection*{Charged Superradiance in de Sitter Space}

We now study the regime of parameters for which black holes in de Sitter space decay by emitting non-superradiant modes, corresponding to $\Delta S_b \le 0$. For completeness, we also derive  the regime for which   $\Delta S_c \leq 0$. As before, we consider the following change of the mass and charge of the black hole during the tunneling process:   $({\cal M},{\cal Q})\to({\cal M}-\mathfrak{m},{\cal Q}-\mathfrak{q})$. We write the horizon radii before emission as $r_{b,c}^i$ and after as $r_{b,c}^f$. Then, the entropy difference of either the black hole or cosmological horizon vanishes, of course, when $r_b^f = r_b^i$ and $r_c^f = r_c^i$, respectively. Using the general method described in  Appendix \ref{app:generalentropybound} we find the compact expressions
\beq
\bal
\label{eq:boundmass}
\Delta S_{b} &\leq 0 : \qquad \mathfrak{m} \geq {\cal M} + \frac{(r^i_b)^3}{\ell_4^2} - \frac{(r^i_b)^2+({\cal Q}-\mathfrak{q})^2}{r^i_b} ~, \\
\Delta S_{c} &\leq 0 : \qquad \mathfrak{m} \leq {\cal M} + \frac{(r^i_c)^3}{\ell_4^2} - \frac{(r^i_c)^2+({\cal Q}-\mathfrak{q})^2}{r^i_c} ~. \\
\eal
\eeq
As explained in Appendix \ref{app:generalentropybound}, the general recipe   can also be used to derive an expression  for $\Delta S_b \leq -C$, where $C$ is a constant independent of $\sqrt{G_4}/\ell_4$. As we will see later, this bound is relevant for the FL bound since   it leads to   suppressed decay. This results in a similar expression as \eqref{eq:boundmass} with the initial horizon radius now replaced by $r_C = \sqrt{(r^i_b)^2- G_4C / \pi}$. We find
\beq
\bal \label{eq:DeltaSCexpressions}
\Delta S_b &\leq -C : \qquad \mathfrak{m} \geq {\cal M} + \frac{(r_C)^3}{\ell_4^2} - \frac{(r_C)^2+({\cal Q}-\mathfrak{q})^2}{r_C} ~.
\eal
\eeq
We will refer to decays obeying $\Delta S_b \leq -1$ as suppressed and decays obeying $\Delta S_b > -1 $ as unsuppressed, because the probability rate of the former decays are exponentially suppressed.  We could have chosen a different negative number instead of $-1$, but we follow the   convention in \cite{Montero:2019ekk}. 

Before we consider the three extremal  limits of these general results, we comment on the proper definition of energy. Just as for charged black holes in flat space   we can define energy with respect to different observers in de Sitter space. In particular, we take the same definition of energy $m_s(r)$ \eqref{Uenergy}, but now with the blackening factor  appropriate for Reissner-Nordstr\"{o}m-de Sitter black holes.   In de Sitter space the natural observer with respect to which the energy is defined is the geodesic observer $r_{\cal O}$, defined by   $f_+'(r_{\cal O})=0$, corresponding to a local maximum of the blackening factor in between the outer black hole and cosmological horizon \cite{Bousso:1996au}. In empty de Sitter space this yields $r_{\cal O} = 0$ and in the Nariai limit $r_{\cal O}=r_b$. Unlike flat space, there is no asymptotic region to define energy in between the black hole and cosmological horizon. Instead, if we consider decays of small black holes to pure de Sitter space, i.e. we evaluate for $(\mathfrak{m},\mathfrak{q})=({\cal M},{\cal Q}) \ll \ell_4 $, and take the limit $r\to 0$ we obtain
\beq \label{eq:r=0energy}
\lim_{r\to0} \left.m_s(r)\right|_{(\mathfrak{m},\mathfrak{q})=({\cal M},{\cal Q}) \ll \ell_4} = \frac{\mathfrak{m}}{2G_4} = m ~.
\eeq
This is the energy defined with respect to a static observer in de Sitter at $r=0$. On the other hand, in the charged Nariai limit we find  
\beq \label{eq:r=rbenergy}
\lim_{r\to r_{b}} m_s(r) =\frac{\sqrt{f_-(r_b)}}{4\pi G_2\ell_{S^2}} ~,
\eeq
Here,  the radius of the $S^2$  coincides with the Nariai black hole horizon radius, $\ell_{S^2}=r_b$. An important difference with respect to black holes in flat space is that for a decay close to the extremal Nariai limit, there no longer is a well-defined observer for which the shell energy is $m$~\eqref{eq:r=0energy}. Instead, we have to use \eqref{eq:r=rbenergy} as our definition of energy, corresponding to the energy of a near-horizon observer.

We now consider the extremal limits of the general bounds \eqref{eq:boundmass} and \eqref{eq:DeltaSCexpressions}. For reasons that will become clear we are mainly interested in the bounds related to the black hole horizon, but the expressions for the entropy difference of the cosmological horizon can also be easily obtained. We     express the  bound \eqref{eq:boundmass} in terms of   $\mathfrak{m}$ and  $\mathfrak{q}$
\beq
\bal \label{eq:DeltaS0expressions}
\text{Cold Black Hole:} \qquad \Delta S_b &\leq 0 \quad \leftrightarrow \quad \mathfrak{m} \geq \frac{\sqrt{6}\mathfrak{q}\left(2{\cal Q}-\mathfrak{q}\right)}{\sqrt{\ell_4^2-\sqrt{\ell_4^4-12\ell_4^2{\cal Q}^2}}} ~, \\
\text{Charged Nariai Black Hole:} \qquad \Delta S_b &\leq 0 \quad \leftrightarrow \quad \mathfrak{m} \geq \frac{\sqrt{6}\mathfrak{q}\left(2{\cal Q}-\mathfrak{q}\right)}{\sqrt{\ell_4^2+\sqrt{\ell_4^4-12\ell_4^2{\cal Q}^2}}} ~, \\
\text{Ultracold Black Hole:} \qquad \Delta S_b &\leq 0 \quad \leftrightarrow \quad  \mathfrak{m} \geq \sqrt{2}\mathfrak{q}\left(1 - \frac{\sqrt{3}}{\ell_4}\mathfrak{q}\right) ~. \\
\eal
\eeq
In terms of the  horizon energy $m_s$ these three bounds simply reduce to $m_s \geq 0$, again indicating that the superradiant regime is decoupled in the near-horizon limit for positive  horizon shell energy. Furthermore, the  bound $\Delta S_b \leq -C$ takes a somewhat complicated form in terms of $m_s$, even in the different extremal limits. Although it is easily obtained explicitly, for ease of presentation we only show it for the ultracold black hole in the limit $\sqrt{G_4}/\ell_4 \ll 1$.   Expressed in ADM units this leads to
\beq \label{eq:DeltaSuc}
\Delta S_b \leq -C \quad \leftrightarrow \quad m_s \geq\left(\frac{\sqrt{6}qHM_pC}{\pi}\right)^{1/2}
\left(1-\frac{\sqrt{\frac{3}{2}} H q}{8 \pi M_p} + {\cal O}\left(H/M_p\right)^2\right)
\eeq
For $C=1$ we recognize the leading term as the exact FL bound, including the numerical factor obtained in \cite{Montero:2021otb}. Thus, we see that in the probe limit $\sqrt{G_4}/\ell_4\ll1$ the FL bound corresponds to suppressed emission $\Delta S_b \leq -1$. The second term is the first correction in $H/M_p$ to the FL bound, which  is new in the literature.

\begin{figure}[t]
\centering
\includegraphics[scale=.75]{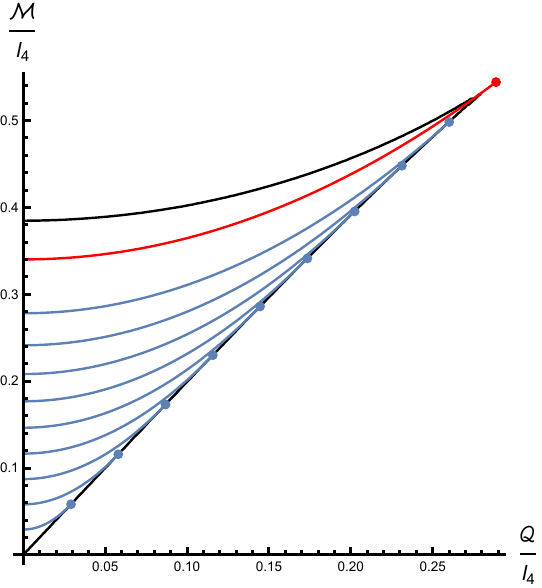}
\caption{The `shark fin' diagram for RNdS black holes. The different colored lines correspond to $\Delta S_b = 0$ where the initial state is a cold black hole with a given charge. The region above a particular line indicates $\Delta S_b > 0$ for that black hole. For  the red line the initial state is the  ultracold black hole. The dots indicate the initial black holes.}
\label{fig:colddecay}
\end{figure}
\begin{figure}[b]
\centering
\includegraphics[scale=.75]{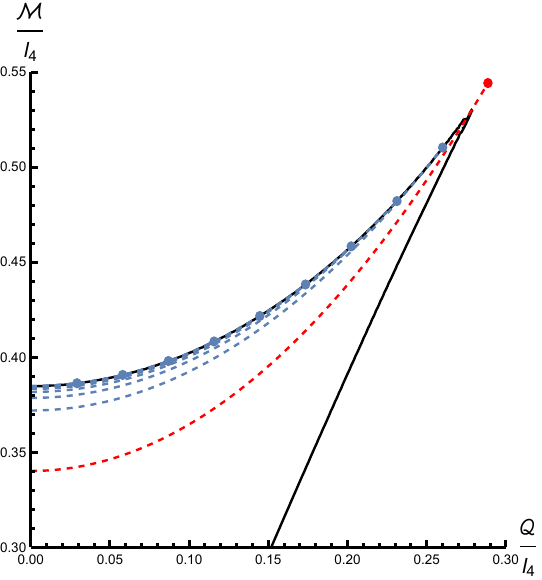}
\caption{Part of the `shark fin' diagram. The different dashed lines correspond to $\Delta S_b = 0$ where the initial state is a Nariai black hole with a given mass. The region above a particular line indicates $\Delta S_b > 0$ for that black hole. For  the red line the initial state is the  ultracold black hole.  The dots indicate the initial black holes.  }
\label{fig:Nariaidecay}
\end{figure}

In Figures \ref{fig:colddecay} and \ref{fig:Nariaidecay} we display the lines of zero entropy difference, $\Delta S_b=0$, indicating the boundary of superradiant decays in the shark fin. Having detailed the boundary between (un)suppressed and (non-)superradiant decays, in the next section we will consider if there are decays that exit the shark fin.

\section{Higher-Order Backreaction and Charged Black Hole Decay} \label{sec:Bounds}
We are now ready to put our derived expressions for the entropy difference to work and see how they relate to potential decay channels of charged black holes in de Sitter space that exit the shark fin. Because empty de Sitter space is a state of maximum entropy, it is reasonable to expect that---from a thermodynamics point of view---all charged black holes decay to empty de Sitter space by following a trajectory within the shark fin consisting of RNdS black hole solutions.

However, as was pointed out in \cite{Montero:2019ekk} such a (mild) decay of charged black holes is not guaranteed in de Sitter space. Close to the charged Nariai branch, an unsuppressed decay channel exists that results in a fast discharge of the black hole without being accompanied by a large decrease in the mass. If this discharge is large enough, this would result in a decay trajectory that exits the shark fin. To prevent this pathological situation, the authors of \cite{Montero:2019ekk} proposed a bound (the FL bound) on the allowed mass and charge of elementary states in order to ensure the absence of such unsuppressed decays.

They derived the FL bound from the fact that the near-horizon decay rate close to the Nariai branch is governed by Schwinger pair production in dS$_2$. By assuming that the stress-energy tensor of emitted particles in the near-horizon Nariai limit takes the form of thermal radiation they argued that---whenever Schwinger pair production is unsuppressed---the resulting geometry describes a big crunch. They  interpreted this as a decay channel that exits the shark fin and hence should be prohibited. Thus, requiring the near-horizon Schwinger rate in the probe limit to be suppressed resulted in the FL bound, which we saw is given by $\Delta S_b \leq -1$ in the probe limit.

Using our results from the tunneling method, which takes into account higher-order backreaction effects, below we show the following statements:
\begin{itemize}
\item There are superradiant decay channels $(\Delta S_b >0)$ that do not exit the shark fin, but have imaginary values of the shell energy $m_s$.
\item Suppressed decays channels ($\Delta S_b \le-1$) and decays that are unsuppressed but not superradiant ($0 \geq\Delta S_b >-1$)  never exit the shark fin and are described by real $m_s$.
\item If   $m_s$ is real, superradiance is decoupled and no decay (suppressed or unsuppressed) exits the shark fin.
\end{itemize}
Especially the third point is relevant for the FL bound, as   the bound is expressed in terms of the energy parameter $m_s$.  As we will explain below, for decays described by the tunneling computation  the third point shows that violating the FL bound does not lead to a big crunch for every observer.

\subsection{Unsuppressed Decays Do Not Exit the Shark Fin}
\begin{figure}[t]
\centering
\includegraphics[scale=.9]{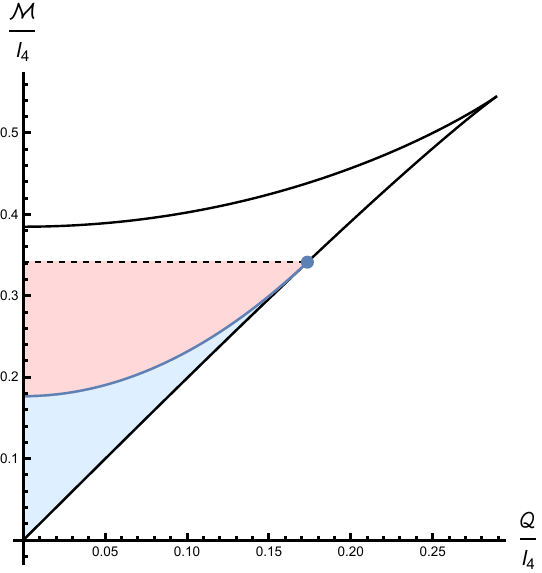}
\caption{The `shark fin' diagram of RNdS black hole solutions. The blue line indicates a decay of a cold black hole along a $\Delta S_b=0$ trajectory. The blue shaded region corresponds to non-superradiant ($\Delta S_b<0$) decay and the red shaded region to superradiant ($\Delta S_b>0$) decay. The dashed line indicates that we are only looking at decays that decrease the mass and charge. The dot corresponds to the initial black hole. Superradiant decays never occur close to the extremal line.}
\label{fig:ColdDecaySingle}
\end{figure}

Let us first comment on charged superradiance for cold black holes far away from the ultracold point and the absence of decay channels that exit the shark fin. We recall our definitions that decays satisfying $\Delta S_b > -1$ are unsuppressed and decays obeying $\Delta S_b >0$ are superradiant. Cold black holes that are small with respect to the de Sitter radius behave in the same way as charged black holes in flat space. As we discussed previously, for such black holes the presence of a superradiant decay channel causes them to loose charge efficiently. Indeed, this can also be seen in Figure \ref{fig:colddecay} and  in more detail in Figure \ref{fig:ColdDecaySingle}. The regime of superradiant decays does not `touch' the extremal cold boundary of the shark fin implying there is no danger that a cold black hole decays to exit the shark fin by emitting a superradiant particle. Instead, superradiant decay channels take a black hole on the cold line to another non-extremal black hole within the shark fin.

\begin{figure}[t]
\centering
\includegraphics[scale=.9]{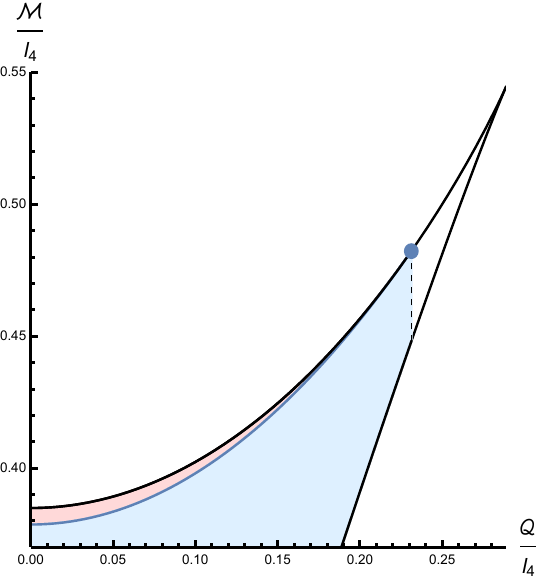}
\caption{The `shark fin' diagram  of RNdS black hole solutions. The blue line indicates a decay of a charged Nariai black hole along a $\Delta S_b=0$ trajectory. The blue shaded region corresponds to non-superradiant ($\Delta S_b<0$) decay and the red shaded region to superradiant ($\Delta S_b>0$) decay. The dashed line indicates that we are only looking at decays that decrease the mass and charge. The dot corresponds to the initial black hole. The superradiant regime is attached to the extremal line.}
\label{fig:NariaiDecaySingle}
\end{figure}

This should be contrasted with the situation for decays from black holes close to the charged Nariai branch. In this case, if we start with a charged Nariai black hole the regime of superradiant decays overlaps with the extremal lime of the charged Nariai branch, see Figure \ref{fig:NariaiDecaySingle}. 
This suggests that---without a bound on the spectrum of allowed particles---there is a potential danger of exiting the shark fin via an unsuppressed process. Whether or not this happens depends on the details of the decay process.  Figure \ref{fig:NariaiDecaySingle} indicates this is a subtle question. The red region corresponds to decays that are superradiant, but nonetheless occur within the shark fin. Thus, it is clear that at least not all superradiant decay channels exit the shark fin. To see this, it is crucial to take into account backreaction. Indeed, for any black hole with final mass ${\cal M}_f$ after emission, the mass needs to obey ${\cal M}_C \leq {\cal M}_f \leq {\cal M}_N$ to remain within the shark fin, where  ${\cal M}_N$ and ${\cal M}_C$ are the mass of a Nariai and cold black hole, respectively, as given by \eqref{eq:NariaiColdMass}. Starting initially with a charged Nariai black hole this bound translates to
\begin{align} \label{eq:NoNaked}
\text{Not Exiting the Shark Fin:} \quad \mathfrak{m} \,\,\geq \,\, &\frac{\ell_4}2\sqrt{\frac23}\Bigg(\sqrt{1+\frac{36{\cal Q}^2}{\ell_4^2}+\left(1-\frac{12{\cal Q}^2}{\ell_4^2}\right)^{3/2}} \\
- &\sqrt{1+\frac{36({\cal Q}-\mathfrak{q})^2}{\ell_4^2} +\left(1-\frac{12({\cal Q}-\mathfrak{q})^2}{\ell_4^2}\right)^{3/2}} \Bigg) ~. \nonumber
\end{align}
 For uncharged radiation $\mathfrak{q}=0$ the bound simplifies to $\mathfrak{m}\geq 0$. Hence, emission of (massless) neutral particles never leads to an exit of the shark fin. At the same time, starting from a charged Nariai black hole the absence of superradiance requires that  (see \eqref{eq:DeltaS0expressions})
\beq \label{eq:NoSR}
\text{No Superradiance:} \quad \mathfrak{m} \geq \frac{\sqrt{6}\mathfrak{q}\left(2{\cal Q}-\mathfrak{q}\right)}{\sqrt{\ell_4^2+\sqrt{\ell_4^4-12\ell_4^2{\cal Q}^2}}} ~.
\eeq
One can show that the right-hand side of this inequality is always larger than the right-hand side of \eqref{eq:NoNaked}. Thus, it is never possible to exit the shark fin from a non-superradiant decay channel. If we violate \eqref{eq:NoSR}, it is possible---but not guaranteed---that we also violate \eqref{eq:NoNaked}. This explains the origin of the red region in Figure \ref{fig:NariaiDecaySingle}, as it corresponds to a superradiant regime which nevertheless lies within the shark fin.

Now consider what happens when we take the probe limit $\sqrt{G_4}/\ell_4 \ll 1$, ignoring backreaction. In that case, the two inequalities become equal:
\beq
\bal \label{eq:NoNakedLimit}
\text{Not Exiting the Sharkfin:} \qquad \mathfrak{m} &\geq \frac{\sqrt{3}\mathfrak{q}}{\ell_4}(2{\cal Q}-\mathfrak{q}) + {\cal O}\left(\frac{G_4}{\ell_4^2}\right)  ~,\\
\text{No Superradiance:} \qquad \mathfrak{m} &\geq \frac{\sqrt{3}\mathfrak{q}}{\ell_4}(2{\cal Q}-\mathfrak{q}) + {\cal O}\left(\frac{G_4}{\ell_4^2}\right) ~.
\eal
\eeq
Thus, if we did not take into account backreaction we would have concluded that all superradiant decays exit the shark fin.

Moreover, in the probe limit we do not only find that the bound on (the absence of) superradiance coincides with the bound on decays that stay within the shark fin. Any bound of the form $\Delta S_b \leq -C$ coincides in the probe limit with the bound on decays that do not exit the shark fin, when $C$ is independent of $\sqrt{G_4}/\ell_4$. This can be seen from   expression~\eqref{eq:DeltaSCexpressions}. Evaluating for the initial state being a charged Nariai black hole and expanding for $\sqrt{G_4}/\ell_4\ll 1$ we find
\beq \label{eq:DeltaSexp}
\Delta S_b \leq -C : \qquad \mathfrak{m} \geq \frac{\sqrt{3}\mathfrak{q}}{\ell_4}(2{\cal Q}-\mathfrak{q}) + {\cal O}\left(\frac{G_4}{\ell_4^2}\right) ~.
\eeq
Hence, to leading order the bound is independent of $C$ and coincides with \eqref{eq:NoNakedLimit}. Again, if we did not take into account backreaction one might have concluded that the tunneling rate needs to be sufficiently suppressed ($\Delta S_b \ll -C$, with $C\geq 1$) to prevent exiting the shark fin.

Up untill now we expressed the relevant bounds in terms of $\mathfrak{m}$. However, as argued before the appropriate energy parameter in the Nariai limit is $m_s(r_b)$ \eqref{eq:r=rbenergy}, evaluated at the black hole horizon, and therefore we express any potential bound in terms of this energy. Indeed, in Sec. \ref{sec:NHlimits} we show that only when expressed in terms of this energy parameter does the decay rate of Nariai black holes reduce to known expressions for Schwinger pair production in dS$_2$. Therefore, we will now express the bounds \eqref{eq:NoNaked} and \eqref{eq:NoSR} in terms of $m_s$. In the Nariai limit, the relation between $\mathfrak{m}$ and the  horizon shell energy   $m_s$ is given by
\beq
\mathfrak{m} = \frac{\sqrt{6}\big((G_4m_s)^2+\mathfrak{q}\left(2{\cal Q}-\mathfrak{q}\right)\big)}{\sqrt{\ell_4^2+\sqrt{\ell_4^4-12\ell_4^2{\cal Q}^2}}} ~.
\eeq
With $m_s=0$ we recognize this as saturating the `No Superradiance' bound \eqref{eq:NoSR}. Because $m_s$ appears quadratically in this expression, any real and non-zero value of $m_s$ results in a decay process that is above the `No Superradiance' bound. This implies that no decay channel involving a real  horizon energy   $m_s$ exits the shark fin. Again, in the probe limit the term proportional to $m_s$ is subdominant and this expression saturates \eqref{eq:NoNakedLimit} and \eqref{eq:DeltaSexp}. Thus, without taking into account backreaction, one might incorrectly conclude that violating a bound of the form $\Delta S_b \leq -C$ exits the shark fin.  We have therefore identified a set of unsuppressed decay channels whose endpoint is still described by   RNdS black holes.

One might wonder if this conclusion changes if the energy $m_s(r)$ is defined with respect to an observer positioned at a different radial location $r=r_{\cal O}\neq r_b$, as appropriate for a non-extremal black hole. We will now show that this is not the case. In general, the geodesic observer is located in between the outer black hole horizon and the cosmological horizon. This location $r_{\cal O}$ corresponds to the local maximum of the blackening factor, where  $f'(r_{\cal O})=0$. We   define the energy with respect to that observer as
\beq
m_s(r_{\cal O}) = 4\pi r_{\cal O}^2{\cal T}(r_{\cal O}) ~,
\eeq
where ${\cal T}(r)$ is defined in \eqref{eq:tensiondef}. We will now argue that there is no critical value of $m_s(r_{\cal O})$ below which a decay exits the shark fin. For $m_s(r_{\cal O})$ to be real it is necessary that both $f_\pm(r_{\cal O}) \geq 0$. The blackening factors are positive in between the outer black hole and cosmological horizon:
\beq
\bal
r_b^i\leq r\leq r^i_c \quad \rightarrow \quad f_+(r)&\geq 0~, \\
r_b^f\leq r\leq r^f_c \quad \rightarrow \quad f_-(r)&\geq 0~.
\eal
\eeq
Here $r^i_{(b,c)}$ refers to the horizon radii before emission and $r^f_{(b,c)}$ to the horizon radii after emission. Thus, real $m_s(r_{\cal O})$ requires that $r_b^{(i,f)} \leq r_{{\cal O}}\leq r_c^{(i,f)} $. Clearly, this inequality cannot be satisfied for decay channels exiting the shark fin, as this implies $r^f_{(b,c)} \in \mathbb{C}$. Hence, any decay channel that exits the shark fin can only be described by imaginary~$m_s(r_{\cal O})$.

\section{Extremal Near-Horizon Limits} \label{sec:NHlimits}
In this section we take the near-horizon limit of the entropy difference $\Delta S_b$ for the   extremal charged black holes in de Sitter space. As we will see, the near-horizon geometries of the different extremal RNdS black holes are given by: AdS$_2\times S^2$, dS$_2\times S^2$ or Mink$_2\times S^2$  \cite{Romans:1991nq,Mann:1995vb,Cardoso:2004uz,Castro:2022cuo}. Reducing over the two-sphere yields an effective two-dimensional spacetime that is unstable to nucleation of a domain wall. The energy of this domain wall is given by $m_s$.

\subsubsection*{Cold Black Hole}
First, we consider the near-horizon limit of the cold black hole geometry. We already saw that extremal electrically charged black holes in flat space have an AdS$_2\times S^2$ near-horizon geometry and this generalizes to the cold black hole, albeit with a different radius. First, we take the limit $r_a\to r_b$ and then perform the coordinate transformation
\beq
r=r_b + \epsilon\, \ell_2^2 \sigma ~, \quad  t = \frac{\tau}{\epsilon} ~,
\eeq
where we defined
\beq \label{eq:AdSradius}
\ell_2 = \frac{\ell_4 r_b}{\sqrt{(r_c-r_b)(r_c-3r_b)}} ~.
\eeq
Zooming in on the horizon by taking $\epsilon \to 0$, we find
\beq
\rmd s^2 = \ell_2^2\left(-\sigma^2\rmd\tau^2 + \frac{\rmd\sigma^2}{\sigma^2}\right) + r_b^2\, \rmd\Omega_2^2 ~.
\eeq
After performing an additional transformation $\sigma\to1/\zeta$ this reduces to the AdS$_2\times S^2$ metric in Poincaré coordinates:
\beq
\rmd s^2 = \frac{\ell_2^2}{\zeta^2}\left(-\rmd\tau^2 + \rmd\zeta^2\right) + r_b^2\, \rmd\Omega_2^2 ~.
\eeq
This identifies $\ell_2$ as the AdS$_2$ radius.

Because the near-horizon geometry is AdS$_2\times S^2$, it is expected that in the probe limit the decay rate should again be given by Schwinger pair production in AdS$_2$. Writing $\mathfrak{m}$ in terms of the near-horizon energy, we find
\beq
\mathfrak{m} = \frac{\sqrt{6}\left(\tilde m_s^2+\mathfrak{q}(2{\cal Q}-\mathfrak{q})\right)}{\left(\ell_4^2-\sqrt{\ell_4^4-12\ell^2{\cal Q}^2}\right)^{1/2}} ~.
\eeq
Here we found it convenient to define $\tilde m_s=G_4m_s$. Plugging this into the general expression for $\Delta S_b$ results in a rather complicated expression.

Using the relation \eqref{eq:AdSradius} for the AdS$_2$ radius, we can express the result in terms of the ratio $\ell_2/\ell_4$. Black holes along the cold line far away from the ultracold point correspond to the limit $\ell_2/\ell_4\to 0$. In this limit the mass reduces to
\beq
\mathfrak{m} = 2\mathfrak{q} - \frac{\mathfrak{q}^2}{\ell_2} + \frac{\tilde m_s^2}{\ell_2} + {\cal O}(\ell_2/\ell_4) ~.
\eeq
This coincides with previously known expressions in flat space \cite{Aalsma:2018qwy}. Taking the same limit in the expression for $\Delta S_b$ and afterwards expanding for large $
\ell_2/\mathfrak{q}$ (a probe limit) we find
\beq
\Delta S_b = -\frac{2\pi \ell_2}{G}\left(\mathfrak{q} - \sqrt{\mathfrak{q}^2-\tilde m_s^2}\right) + {\cal O}(\mathfrak{q}/\ell_2) + {\cal O}(\ell_2/\ell_4) ~.
\eeq
As is well known, Schwinger pair production in AdS requires a minimum charge of $\mathfrak{q}\geq \tilde m_s$. In four-dimensional quantities, this corresponds to $\mathfrak{m}\leq 2\mathfrak{q}$ which is the Weak Gravity Conjecture in flat space.

Restoring `ADM units' using \eqref{eq:masschargepara} we can write the decay rate associated to Schwinger pair production as
\beq
\Gamma \sim \exp\left(-2\pi \ell_2^2\left(qE - \sqrt{(qE)^2 - (m_s/\ell_2)^2}\right)\right) ~,
\eeq
which agrees with the known expression for AdS$_2$ in the literature \cite{Pioline:2005pf,Kim:2008xv}.

\subsubsection*{Nariai Black Hole}
To obtain the near-horizon limit of the (charged) Nariai black hole we first take the limit $r_b\to r_c$ and perform a similar coordinate transformation
\beq
r=r_c - \epsilon\, \ell_2^2 \tau ~, \quad  t = \frac{\sigma}{\epsilon} ~,
\eeq
where
\beq \label{eq:dS2radius}
\ell_2 = \frac{\ell_4 r_c}{\sqrt{(r_c-r_a)(r_a+3r_c)}} ~.
\eeq
This yields
\beq
\rmd s^2 = \ell_2^2\left(-\frac{\rmd\tau^2}{\tau^2} + \tau^2\rmd\sigma^2\right) + r_c^2\, \rmd\Omega_2^2 ~.
\eeq
Performing the transformation
\beq
\sigma = x ~, \quad \tau = 1/\eta ~,
\eeq
we recognize this as the dS$_2\times S^2$ metric in planar coordinates:
\beq
\rmd s^2 = \frac{\ell_2^2}{\eta^2}\left(-\rmd\eta^2+\rmd x^2\right) +r_c^2\rmd\Omega_2^2 ~.
\eeq
Expressing the dS$_2$ part in static coordinates, we can use the timelike Killing vector
\beq
\zeta = \partial_{t_{\rm static}} = -\frac\eta{\ell_2}\partial_\eta -\frac x{\ell_2}\partial_x ~,
\eeq
to define the surface gravity. At the location of the (future) dS$_2$ horizon, given by $\eta+x = 0$ this leads to a temperature of
\beq
T = \left.\frac{\kappa}{2\pi}\right|_{\eta+x=0} = \frac1{2\pi\ell_2} ~.
\eeq
From the expression of the two-dimensional de Sitter radius \eqref{eq:dS2radius}, we notice that the temperature goes to zero in the limit $r_a\to r_c$, which corresponds to the ultracold black hole. This signals a change in the near-horizon geometry to Minkowski space.

Because Nariai black holes have a finite temperature (away from the ultracold point) we expect emission in the near-horizon limit to be governed by a mixture of Hawking radiation and Schwinger pair production. We evaluate the decay rate for two different Nariai limits: close to the uncharged Nariai limit and at the intersection of the Nariai and so-called lukewarm line (for which the black hole horizon and cosmological horizon have the same temperature).

Far away from the ultracold point we reach the uncharged Nariai black hole, where the two-dimensional de Sitter radius is $\ell_2 = \ell_4/\sqrt{3}$ and the mass in terms of the  horizon energy is given by
\beq
\mathfrak{m} = \frac{\tilde m_s^2-\mathfrak{q}^2}{\ell_2} ~.
\eeq
Taking the probe limit $\ell_2/\mathfrak{q} \gg 1$, the entropy difference is
\beq
\Delta S_b = -\frac{2\pi\ell_2 \tilde m_s}{G_4} + {\cal O}\left(\mathfrak{q}/\ell_2\right)^0 ~.
\eeq
Expressing this in terms of $m_s$, it agrees with   the Schwinger pair production rate of uncharged radiation in dS$_2$ \cite{Kim:2008xv,Frob:2014zka}:
\beq
\Gamma \sim \exp\left(-2\pi\ell_2 m_s\right) ~.
\eeq
We recognize this as the thermal Boltzmann factor in two-dimensional de Sitter space.

At the intersection of the Nariai and lukewarm line the two-dimensional de Sitter radius is given by $\ell_2 = \ell_4/\sqrt{2}$. The mass in terms of the  horizon energy is given by
\beq
\mathfrak{m} = \mathfrak{q} + \frac{\sqrt{2}}{\ell_2}\left(\tilde m_s^2-\mathfrak{q}^2\right) ~.
\eeq
Again taking the probe limit $\ell_2/\mathfrak{q} \gg 1$, the entropy difference becomes
\beq
\Delta S_b = \frac{\sqrt{2}\pi\ell_2 }{G_4}\left(\mathfrak{q}-\sqrt{2m_s^2+\mathfrak{q}^2}\right) + {\cal O}\left(\mathfrak{q}/\ell_2\right)^0 ~.
\eeq
In terms of the electric field at the intersection of the Nariai and lukewarm line, given by
\beq
E = \frac1{\sqrt{\pi G_4}\ell_2^2} ~,
\eeq
the decay rate becomes
\beq \label{eq:dS2decayrate}
\Gamma \sim \exp\left(2\pi\ell_2^2\left(qE-\sqrt{(m_s/\ell_2)^2+(qE)^2}\right)\right) ~.
\eeq
This coincides  with the Schwinger pair production rate in dS$_2$ \cite{Kim:2008xv}.  We can also compare with reference \cite{Frob:2014zka} which computes the exact Schwinger rate, not just the exponential dependence. The exponential dependence \eqref{eq:dS2decayrate} coincides with the exponential dependence for `downward' tunneling \cite{Frob:2014zka} finds. In order to match their exact (but non-backreacted) result requires also taking into account the process of `upward' tunneling in addition to computing the one-loop determinant in the tunneling amplitude. Since the contribution of `upward' tunneling to the Schwinger current is always smaller than the contribution of `downward' tunneling (in the regime $\Delta S_b < 0$) we chose to ignore the upward process. An important difference with pair production in the AdS near-horizon region is that in de Sitter, no minimum charge is required.

\subsubsection*{Ultracold Black Hole}
Lastly, we consider the ultracold black hole. We first take the limit $r_a\to r_b\to r_c$ and perform the coordinate transformation
\beq
r = r_c - \epsilon  + \sqrt{\frac{4\epsilon^3}{r_c\ell_4^2}}X ~, \quad t = \sqrt{\frac{r_c\ell_4^2}{4\epsilon^3}}T ~.
\eeq
Taking the near-horizon limit by sending $\epsilon\to 0$ the metric becomes  
\beq
\rmd s^2 = -\rmd T^2 + \rmd X^2 + r_c^2\rmd\Omega_2^2 ~.
\eeq
This geometry is $\text{Mink}_2\times S^2$ and has a vanishing temperature. We now expect the tunneling rate to be given by Schwinger pair production in flat space.

The  horizon energy is defined via
\beq
\mathfrak{m} = \sqrt{2}\mathfrak{q} - \frac{\sqrt{6}\left(\mathfrak{q}^2-\tilde m_s^2\right)}{\ell_4} ~.
\eeq 
If we now express the entropy difference in terms of   $\tilde m_s$ and take the limit of large four-dimensional de Sitter radius, we find the following result
\beq
\Delta S_b = -\frac{4\pi \ell_4 \tilde m_s^2}{\sqrt{3}G_4 \mathfrak{q}} + {\cal O}(\mathfrak{q}/\ell_4)^0 ~.
\eeq
The electric field of the ultracold black hole is given by
\beq
E = \frac{\sqrt{3}}{\sqrt{4\pi G}\ell_4} ~.
\eeq
Thus, the decay rate becomes
\beq \label{eq:ultracolddecay}
\Gamma \sim \exp\left(-\frac{\pi m_s^2}{Eq}\right) ~.
\eeq
Unsurprisingly, given that the near-horizon region is two-dimensional Minkowksi space, this reduces to the leading contribution to the Schwinger pair production rate in  flat space.

\section{Conclusion} \label{sec:Conclusion}
In this paper we studied the subleading backreaction corrections to charged black hole decay in flat space and de Sitter space using a tunneling approach. This method computes the emission of spherically symmetric, charged radiation from black holes by viewing this decay as a tunneling process. This yields a universal decay probability--taking into account backreaction---that depends on the exponential of the entropy difference of the black hole before and after emission. Charged decays come in a few flavors: suppressed decays that obey $\Delta S_b \leq -C$ with $C\geq 1$ and unsuppressed decays that obey $\Delta S_b > -C$. We make a further distinction between decays that are unsuppressed, but not superradiant ($-C <  \Delta S_b \leq 0$) and superradiant decays ($\Delta S_b > 0$).

By studying the decay of charged Nariai black holes we found that, when backreaction is included, not all superradiant decays lead to naked singularities. Moreover, in the near-horizon limit it is appropriate to describe decays in terms of a  horizon   energy parameter $m_s$. When this parameter is real, the superradiant regime is completely decoupled such that for decays obeying $m_s^2\geq 0$ there is no danger of exiting the sharkfin.

Our results lead to new insights into the FL bound proposed in \cite{Montero:2019ekk}, because  we found this bound is naturally expressed in terms of $m_s$. The FL bound follows from requiring the dS$_2$ Schwinger rate in the near-horizon Nariai limit to be exponentially suppressed. The authors of \cite{Montero:2019ekk} derived that the exponential part of the Schwinger rate in the near-horizon dS$_2$ takes the from
\beq
\Gamma \sim \exp\left(-\frac{\pi m_s^2}{Eq}\right) ~,
\eeq
which we recognize as the $\ell_2/\mathfrak{q}\to\infty$ limit of the expression \eqref{eq:dS2decayrate} or, equivalently, the decay rate in the ultracold point. Here $E$ is the electric field of the ultracold black hole in the near-horizon limit. Requiring the exponent to be suppressed results in the bound (expressed in terms of the Hubble parameter and Planck mass)
\beq \label{eq:ExactFL}
m_s^2 \geq \frac{\sqrt{6}HM_pq}{\pi} ~,
\eeq
which is exactly the bound $\Delta S_b \leq -1 $ evaluated on the ultracold black hole with $C=1$ in the limit $\sqrt{G_4}/\ell_4 \ll 1$ (see \eqref{eq:DeltaSuc}). This matches the exact FL bound including the numerical factor of $\sqrt{6}$.\footnote{The fact that the $\sqrt{6}$ arises in the ultracold point was already observed in \cite{Montero:2021otb}. The original derivation of the Festina Lente bound, however, was performed for the near-horizon dS$_2$ region \cite{Montero:2019ekk}. Here we have obtained this bound  from a non-perturbative tunneling calculation.} This exact agreement shows that the mass appearing in the FL bound is given by our  horizon energy   $m_s$. 

It is possible that the unsuppressed decay process in \cite{Montero:2019ekk}---which involved a multi-particle decay resulting in radiation---is not describable in terms of a tunneling approach. If this is the case, it is necessary to understand whether the multi-particle decay channel in \cite{Montero:2019ekk} or the single-shell decay channel in the tunneling approach is dominant. The argument of \cite{Montero:2019ekk} relies on unsuppressed multi-particle decay modes coming from pair production everywhere in spacetime. The local pair-production action $\Delta S_b$ becomes very small when the FL bound is violated, entering the unsuppressed multi-particle regime. We did not consider this decay mode in this paper, which focuses on the emission of a single shell from the horizon. A full analysis would require to determine whether the multi-particle decay channel in \cite{Montero:2019ekk} or the single-particle decay channel in the tunneling approach is more probable. Our decay probability goes to one in the limit $m_s\to 0$ and $\Delta S_b$ goes to zero in that limit (although this enters the multi-particle regime, where our expressions cease to be valid), so we believe comparing the two decay channels requires further study.

 Assuming the dominance of our single-particle decay channel we can compare with the conclusion of \cite{Montero:2019ekk} that unsuppressed decays lead to a big crunch. Taking into account higher-order backreaction corrections we saw that decays violating $\Delta S_b \leq -1$ do not exit the shark fin. In \cite{Montero:2019ekk}  the authors derived the formation of a big crunch in the near-horizon geometry of the Nariai black hole from the assumption that the stress-energy tensor after discharge of the electric field takes the form of   radiation. In the decay channel that we identify, the big crunch is also present and essentially corresponds to the interior of the black hole, but the endpoint of the decay still represents a  standard RNdS solution. Specifically, taking into account higher-order backreaction corrections we find that for the decay channel described by the tunneling computation it is not guaranteed that a near-horizon observer ends up in the big crunch region. For instance, consider an instantaneous discharge of a charged Nariai black hole   via massless modes $m_s=0$. The endpoint of this decay obeys $\Delta S_b = 0$ and results in a black hole solution with zero charge and  a mass smaller than that of an uncharged   Nariai black hole. Thus, there still is a region in between the two horizons where an observer can avoid witnessing a crunch. This shows that unsuppressed decay channels violating the FL bound exist that, surprisingly, do not lead to a pathology. It would be interesting to understand how the decay channel we studied using the tunneling approach and the one studied in \cite{Montero:2019ekk} are related, if at all. This could be done, for example, by further analyzing the formation of a big crunch singularity due to unsuppressed decay in a four-dimensional setup without imposing a dS$_2\times S^2$ ansatz.

We emphasize  that our conclusion that decays violating $\Delta S_b \le -1$ do not exit the shark fin  strongly relies on the identification of $m_s$ as the appropriate energy parameter in the near-horizon limit. As   already mentioned, this is supported by the exact agreement of the bound $\Delta S_b \leq -1$ with the FL bound when we take the probe limit. Still, one might wonder about the implications of a bound in terms of the mass parameter~$m$, or another energy parameter. We already found that any definition of energy in terms of $m_s(r)$ evaluated for a different  observer excludes decays that exit the shark fin for real~$m_s(r)$. The bound on  $m$ to avoid exiting the shark fin was   stated in~\eqref{eq:NoNaked}, but its validity is ruled out by observations. Considering an ultracold black hole emitting an electron, i.e. $q =\sqrt{4\pi \alpha} \simeq 0.303 $ (where $\alpha$ is the fine-structure constant), this evaluates to $m \gtrsim 10^{26} \rm eV$ in our universe. Such a bound is violated by an electron by 21 orders of magnitude and is therefore ruled out. This does not exclude an interesting bound in terms of a completely different energy parameter.

Another possible caveat is the validity of tunneling computation itself. For example, we did not compute the one-loop determinant multiplying the exponential decay probability and we assumed that decays are effectively described by a single shell. We do not expect the one-loop determinant to modify our conclusions although it would be useful to compute it explicitly. 

Finally, it would be interesting to understand if our results are supported by a complementary analysis. For example, the thermodynamic methods of \cite{Sorce:2017dst} including second-order variations could be a good approach.  It would be particularly illuminating to see if our definition of mass also plays a similar role in a such an analysis. We hope that our work leads to a renewed interest in the constraints imposed by the decay of extremal black holes in de Sitter space.

\subsection*{Acknowledgements}
We would like to thank Miguel Montero, Thomas Van Riet, and Gerben Venken for useful discussions about the interpretation of the FL bound. LA likes to thank the hospitality of the UW-Madison High-Energy Theory group, where part of this work was completed, and Yoshihiko Abe, Puxin Lin and Gary Shiu for useful discussions on related topics. This work benefited from discussions arising at the DAMTP workshop ``Quantum de Sitter Universe'', funded by the Gravity Theory Trust and the Centre for Theoretical Cosmology. This work is part of the Delta ITP consortium, a program of the Netherlands Organisation for Scientific Research (NWO) that is funded by the Dutch Ministry of Education, Culture and Science (OCW). LA is supported by the Heising-Simons Foundation under the “Observational Signatures of Quantum Gravity” collaboration grant 2021-2818. MRV is supported by SNF Postdoc Mobility grant P500PT-206877 “Semi-classical thermodynamics of black holes and the information paradox”.

\appendix

\section{Derivation of General Entropy Bounds} \label{app:generalentropybound}
In this Appendix  we explain the general strategy for deriving the different entropy bounds   used in the main body of this paper. We work in general spacetime dimensions $D$ in this appendix, for which  the blackening factor of Reissner-Nordstr\"{o}m-de Sitter black holes   is  given by
\beq
\label{blackeningD}
f(r) = 1 - \frac{r^2}{\ell_D^2} - \frac{{\cal M}}{r^{D-3}} + \frac{{\cal Q}^2}{r^{2(D-3)}}\,.
\eeq
Here $\ell_D$ is the $D$-dimensional de Sitter curvature radius. We are interested in deriving entropy differences after a decay process where the mass and charge of the black hole change as
\beq
({\cal M},{\cal Q})\to({\cal M}-\mathfrak{m},{\cal Q}-\mathfrak{q}) ~.
\eeq
We use $f_+(r)$ to denote the blackening factor outside the shell and $f_-(r)$ the blackening factor inside the shell. We want to derive   the location where
\beq
\label{deltaSc}
\Delta S_{(b,c)} = -C ~.
\eeq
Defining the radius 
\begin{equation}
    r_C^{D-2}:= (r_{b,c}^i)^{D-2}-\frac{4G_DC}{\Omega_{D-2}}\,,
\end{equation}
where $\Omega_{D-2}= 2\pi^{\frac{D-1}{2}}/\Gamma\left(\frac{D-1}{2}\right)$ is the volume of a unit $S^{D-2}$, equation \eqref{deltaSc} is solved by
\beq
(r_{(b,c)}^f)^{D-2} - (r_C)^{D-2} = 0  \qquad \longrightarrow \qquad r_{(b,c)}^f = r_C \,.
\eeq
By definition, the blackening factor after emission obeys $f_-(r_{(b,c)}^f)=0$, hence   acting with it on both sides of the equation above yields
\beq\label{eq:blackzero}
f_-(r_C)   = 0 ~.
\eeq
By using the explicit form of the blackening factor and specifying $r_C$ we can derive all entropy bounds   in the main body in the paper. 
 Using \eqref{blackeningD} and   \eqref{eq:blackzero} we   find that the bound $\Delta S_b \leq -C$ is equivalent to
\beq\label{eq:blackzeroinm}
\mathfrak{m} \geq {\cal M} - (r_C)^{D-3} \left ( 1- \frac{(r_C)^{2}}{\ell_D^2} \right) - \frac{ ({\cal Q}-\mathfrak{q})^2}{(r_C)^{D-3}} ~. \\
\eeq
For $D=4$ this reduces to the bound \eqref{eq:DeltaSCexpressions} in the paper.

\subsubsection*{Superradiance in Higher Dimensions}

We now use this method to derive the condition for the (absence of) superradiance in higher dimensions. This amounts to imposing $\Delta S_b \leq 0$ and taking $r_C = r^i_b$. This results in the bound
\beq
\Delta S_{b} \leq 0 : \qquad \mathfrak{m} \geq {\cal M} - (r^i_b)^{D-3} \left ( 1- \frac{(r^i_b)^{2}}{\ell_D^2} \right) - \frac{ ({\cal Q}-\mathfrak{q})^2}{(r^i_b)^{D-3}} ~, \\
\eeq
Next we are interested in the special cases of this bound where the initial black hole is extremal. The  mass and charge in the extremal cases are given by \cite{Cardoso:2004uz,Morvan:2022aon} 
\beq
\bal
{\cal M}_{C,N} &= 2 \left ( 1- \frac{D-2}{D-3} \frac{r_{C,N}^2}{\ell_D^2} \right) r_{C,N}^{D-3}, \qquad {\cal Q}_{C,N}^2 =  \left ( 1- \frac{D-1}{D-3} \frac{r_{C,N}^2}{\ell_D^2} \right) r_{C,N}^{2(D-3)}\,,\\
{\cal M}_U &= 4\frac{r_U^{D-3}}{D -1}\, , \qquad {\cal Q}_U = \frac{r_U^{D-3}}{\sqrt{D-2}}\,, \qquad r_U  = \frac{D-3}{\sqrt{(D-1)(D-2)}} \ell_D\,  .
\eal
\eeq
This yields the bounds
\beq
\bal
\text{Cold/Nariai Black Hole:} \qquad  &\mathfrak{m} \geq \frac{\mathfrak{q}}{r_{C,N}^{D-3}}   \left (2 \sqrt{1- \frac{D-1}{D-3}\frac{r_{C,N}^2}{\ell_D^2
}} r_{C,N}^{D-3}-  \mathfrak{q}\right) =  \frac{\mathfrak{q}}{r_{C,N}^{D-3}({\cal Q})}   \left (2{\cal Q} -  \mathfrak{q}\right)~, \\
\text{Ultracold Black Hole:}\qquad  &\mathfrak{m} \geq \frac{2}{\sqrt{D-2}}\mathfrak{q}\left(1 - \frac{\sqrt{D-2}}{2} \left [  (D-3)\sqrt{(D-1)(D-2)} \ell_D\right]^{D-3}  \mathfrak{q}\right) ~. \\
\eal
\eeq

 \subsubsection*{Ultracold Decays in Higher Dimensions}
Next we   evaluate the  bound $\Delta S_b \leq -C$ for the case that the initial state is an ultracold black hole. We express the bound in terms of $m_s$ defined by integrating the tension over the $S^{D-2}$ sphere 
\beq
m_s = \Omega_{D-2}r^{D-2} {\cal T}(r)~.
\eeq
The tension ${\cal T}$ in general $D$ takes the form:
\beq
{\cal T}(r)  = \frac{D-2}{8\pi G_D r}\left[\sqrt{f_-(r)}-\sqrt{f_+(r)}\right] ~. \\
\eeq
For the case where   ultracold black holes are initial states  the  explicit bound $\Delta S_b \leq -C$ looks rather complicated, so we just give the expression for dimensions $D=(4,5,6)$ expanded for $G_D^\frac{1}{D-2}/\ell_D \ll 1$:
\beq
\bal
\Delta S_b \leq -C \quad \leftrightarrow \quad D=4:\quad m_s &\geq \left(\frac{\sqrt{3}C\mathfrak{q}}{G_4\pi\ell_4}\right)^{1/2} + {\cal O}\left(\sqrt{G_4}/\ell_4\right)~, \\
\qquad D=5:\quad m_s &\geq \left(\frac{3C\mathfrak{q}}{2G_5\ell_5}\right)^{1/2} + {\cal O}\left(G_5^{1/3}/\ell_5\right)~, \\
\qquad D=6:\quad m_s &\geq \left(\frac{4\sqrt{5}C\mathfrak{q}}{3G_6\ell_6}\right)^{1/2} + {\cal O}\left(G_6^{1/4}/\ell_6\right)~. \\
\eal
\eeq
These are the FL bounds in higher dimensions. The bounds for other dimensions can be easily obtained using the general method of this appendix.

\bibliographystyle{utphys}
\bibliography{refs}

\end{document}